\documentstyle[11pt,epsf]{article}
\begin{document}
\title{On solitary waves in classical anisotropic Heisenberg chains\\
       with generalized boundary conditions}
\author{John Schliemann and Franz G. Mertens\\
{\it Physikalisches Institut, Universit{\"a}t Bayreuth, D-95440 Bayreuth,
Germany}}
\date{December 1997}
\maketitle
\begin{abstract} 
We examine solitary waves in classical ferromagnetic Heisenberg 
chains with an uniaxial anisotropy and a parallel magnetic field in a 
continuum approach.
The boundary conditions commonly used are generalized to nonlinear
spin wave states, which themselves turn out to be stable only for an anisotropy
of the easy--plane type. In this case we obtain two different branches of
one--soliton--solutions which can be mapped onto each other by a formal
time inversion. Moreover, they show some remarkable similarity to dark
solitons of the Nonlinear Schr\"odinger equation. Numerical simulations for 
the discrete Heisenberg chain show that these solitary waves are highly, 
but not absolutely stable under interaction with linear excitations and as 
well under scattering with each other. The possible significance of these 
solitary waves in a phenomenological theory of one--dimensional magnets is 
briefly addressed.
\end{abstract}


\section{The model}

\label{themodel}
The discrete classical ferromagnetic Heisenberg chain with a local 
uniaxial anisotropy and a magnetic field parallel to this axis is given by
\begin{equation}
\tilde{\cal H}=-\sum_{n}\left[\vec S_{n}\vec S_{n+1}+hS_{n}^{z}
+\frac{\alpha}{2}\left(S_{n}^{z}\right)^2\right]\,.
\label{defmoda}
\end{equation}
The classical spins are unit vectors
$\vec S_{n}=(\sin\vartheta_{n}\cos\varphi_{n},\sin\vartheta_{n}\sin\varphi_{n},
\cos\vartheta_{n})$, $h$ the magnetic field and $\alpha$ the anisotropy
parameter. If we choose the infinite chain to lie in the $x$--direction and
an appropriate length unit with lattice spacing $a=1$, the continuum
approximation of (\ref{defmoda}) in lowest order reads  
\begin{equation}
{\cal H}=\int_{\infty}^{\infty} dx\left(\frac{1}{2}\left(
\frac{\left(\partial_{x}p\right)^{2}}{1-p^2}
+\left(1-p^2\right)
\left(\partial_{x}q\right)^{2}\right)-hp-
\frac{\alpha}{2}p^{2}\right)\,.
\label{defmod}
\end{equation}
Here the classical spins are described by the canonical 
conjugate fields $p(x,t)=\cos(\vartheta(x,t))$,
$q(x,t)=\varphi(x,t)$. The equations of motion are
\begin{eqnarray}
\partial_{t}q&=&-\frac{1}{1-p^2}\partial_{x}^{2}p
-\frac{p}{(1-p^2)^2}(\partial_{x}p)^2-p(\partial_{x}q)^2-h-\alpha p\,,
\label{eom1}\\
\partial_{t}p&=&(1-p^2)\partial_{x}^{2}q
-2p(\partial_{x}p)\partial_{x}q\,.
\label{eom2}
\end{eqnarray}
For $\alpha<0$ and $|\frac{h}{\alpha}|<1$ the solution of lowest energy 
is given by constant $q$ and $p=-\frac{h}{\alpha}$ (easy--plane or, for
finite $h$, easy--cone model), for $\alpha>0$ by $|p|=1$ (easy--axis model).
These field configurations can be used as boundary conditions for the above
equations. In the literature only these two cases appear to be considered,
for a review see \cite{MiSt:91,KIK:90}.\\
Takhtajan \cite{Tak:77} and Fogedby \cite{Fog:80} have shown that the isotropic
model ($\alpha=0$) allows for a Lax representation and is solvable by the
inverse scattering method. Moreover, this model is integrable in the sense
that it has an infinite series of independent conserved quantities. These
results have been extended by Sklyanin \cite{Skl:79} to the case of a general
biaxial anisotropy. The relationship of the model (\ref{defmod}) to the
Nonlinear Schr{\"o}dinger equation has also been established for the
isotropic \cite{ZaTa:79,Lak:77} and the anisotropic case 
\cite{QuCa:82,NaSa:82}. However, all these results have been obtained using
boundary conditions of the above type.  
In the present work we examine the model for more general boundary conditions 
to be specified below.\\
If the spin configuratiuon is the same at both boundaries,
the system has (at least) two well-known conserved quantities apart from the 
energy, namely the total momentum 
\begin{equation}
P=\int dx\,p\partial_{x}q
\label{mom0}
\end{equation}
and the $z$--component of the total angular momentum or magnetic moment 
\begin{equation}
M=\int dx\,p\,.
\label{ang0}
\end{equation}
$P$ is the generator of translations and $M$ is the generator of uniform
rotations of the spins. The Poisson brackets of $\cal H$, $P$, $M$
with each other vanish
since surface terms do not contribute under the above condition. 


\section{Nonlinear spin waves and linear magnons}

From the above equations one easily finds the solutions
\begin{equation}
p=u\quad,\quad q=q_{0}+kx-\omega t
\label{sw}
\end{equation}
with constants $u$, $k$, $q_{0}$, $|u|<1$, and $\omega$ given by
\begin{equation}
\omega=\left(k^{2}+\alpha\right)u+h\,.
\label{dis}
\end{equation}
Here the spin field describes a propagating wave characterized by the 
parameters $u$ and $k$. (\ref{sw}) is an exact solution of the nonlinear 
equations of motion and will therefore be called a nonlinear spin wave.
It has a homogenous energy density obtained 
from (\ref{defmod}) as
\begin{equation}
\varepsilon=\frac{1}{2}\left(1-u^{2}\right)k^{2}
-hu-\frac{\alpha}{2}u^{2}\,.
\end{equation}
Note that the spin configurations mentioned in the previous section
are included as special cases. The time--dependent field configuration
(\ref{sw}) can be used as a generalization of the boundary conditions 
given above. In the next section we will obtain solitary waves
obeying such boundary conditions, i.~e. having the asymptotic form
of a spin wave for $|x|\to\infty$. As the spin waves in general do not 
give absolute minima of the Hamiltonian their stability is to be questioned. 
Therefore let us briefly discuss the stability of the homogenous
spin wave solution (\ref{sw}) if the system is forced to have the 
asymptotic structure
\begin{equation}
\lim_{|x|\to\infty}p=u\quad,\quad\lim_{|x|\to\infty}\partial_{x}q=k\,.
\label{bc}
\end{equation}
The ansatz
\begin{equation}
p(x,t)=u+\eta(x,t)\quad,\quad q(x,t)=q_{0}+kx-\omega t+\xi(x,t)
\end{equation}
with $\eta$, $\xi$ vanishing for $|x|\to\infty$ describes
small fluctuations around the spin wave solution respecting
the boundary condition and the conservation of $P$ and $M$.
Introducing the Fourier transforms
\begin{eqnarray}
\eta(x,t)&=&\frac{1}{\sqrt{2\pi}}\int dl\,\tilde\eta(l,t)e^{\imath lx}\,,\\
\xi(x,t)&=&\frac{1}{\sqrt{2\pi}}\int dl\,\tilde\xi(l,t)e^{\imath lx}
\end{eqnarray} 
this leads to the restriction $\tilde\eta(0,t)=0$. Inserting these 
expressions in the equations of motion and linearizing in $\eta$, $\xi$
gives
\begin{eqnarray}
\partial_{t}\tilde\eta&=&
-2\imath ukl\tilde\eta-l^{2}\left(1-u^{2}\right)\tilde\xi\,,\\
\partial_{t}\tilde\xi&=&\left(\frac{l^2}{1-u^2}-k^{2}-\alpha\right)\tilde\eta
-2\imath ukl\tilde\xi\,.
\end{eqnarray}
The general solution of these equations has the form
$\tilde\eta(l,t)=\tilde\eta_{0}(l)e^{-\imath\omega_{l}t}$,
$\tilde\xi(l,t)=\tilde\xi_{0}(l)e^{-\imath\omega_{l}t}$ with
\begin{equation}
\omega_{l}=2ukl\pm\sqrt{l^{4}+l^{2}\left(1-u^{2}\right)
\left(-\alpha-k^{2}\right)}\,.
\label{ms}
\end{equation}
The nonlinear spin wave solution is stable if $\omega_{l}$ is real for
every Fourier mode $l$. This is the case either for $|u|=1$, or for
\begin{equation}
-\alpha-k^2\geq 0
\label{stab}
\end{equation}
with general $u$. The relation (\ref{stab}) can only be fullfilled for
finite $k$ if $\alpha<0$, and stable exact spin wave solutions 
(\ref{sw}) are restricted to such type of Hamiltonian. If (\ref{stab}) is
valid eq.~(\ref{ms}) is the spectrum of linear excitations above the
nonlinear spin wave. We shall call these excitations magnons.
The extremal group velocities within the two magnon dispersion branches
are achieved at $l=0$ with
\begin{equation}
\left(\frac{d\omega_{l}}{dl}\right)_{l=0}=
2uk\pm\sqrt{\left(1-u^{2}\right)\left(-\alpha-k^{2}\right)}\,.
\label{mgv}
\end{equation} 
The upper sign corresponds to a minimum, the lower to a maximum of
$d\omega_{l}/dl$ on the particular branch.


\section{Solitary waves}

As the nonlinear spin waves are stable uniform solutions under the condition
(\ref{stab}) we use them as a reference state to construct
localized solutions.
We are looking for solitary solutions which have the form of a spin 
wave for $|x|\to\infty$ determined by parameters $u$,~$k$.\\
Inserting the ansatz
\begin{equation}
p(x,t)=p(x-vt)\quad,\quad q(x,t)=\Omega t+\bar q(x-vt)
\label{ansatz}
\end{equation}
and performing two formal integrations one ends up with
(cf.~Fogedby \cite{Fog:80})
\begin{eqnarray}
\frac{d\bar q}{dx} & = & v\frac{r-p}{1-p^2}\,,
\label{eom3}\\
\left(\frac{dp}{dx}\right)^{2} & = & -v^{2}\left(1+r^{2}-2rp\right)
+2\left(h+\Omega\right)p\left(p^{2}-1\right)-s\left(p^{2}-1\right)
+\alpha p^{2}\left(p^{2}-1\right)\,,
\label{eom4}
\end{eqnarray}
where $r$,~$s$ are integration constants, $v$ the velocity of the solitary wave
and $\Omega$ an internal frequency. Denoting the r.h.s. of (\ref{eom4}) by
$F(p)$ we conclude that this quantity should be nonnegative for certain
values of $p$ between $(-1)$ and 1. A real single root of $F(p)$ corresponds
to an extremal value of $p(x)$, while a real double root $p=u$ leads
to a fixed point since one can easily show that in this case also the second
and successively all higher derivatives of $p$ vanish for $p=u$. For our
solution to be localized and asymptotically in the state $p=u$ we need the 
latter case and make the general ansatz
\begin{equation}
\left(\frac{dp}{dx}\right)^{2}=\alpha(p-u)^{4}+\beta(p-u)^3+\gamma(p-u)^2\,.
\label{eom5}
\end{equation} 
To fullfill the additional boundary condition for $q$ we must also have
\begin{equation}
\lim_{|x|\to\infty}\partial_{x}\bar q=v\frac{r-u}{1-u^2}=k\,.
\label{eom6}
\end{equation}
A comparision of eqs.~(\ref{eom3}),~(\ref{eom4}) and (\ref{eom5}),~(\ref{eom6})
leads to
\begin{eqnarray}
vr&=&k\left(1-u^2\right)+uv\,,\\
s&=&v^{2}+2kuv+k^{2}-3u^{2}k^{2}-\alpha u^{2}
\end{eqnarray}
and
\begin{eqnarray}
\beta&=&-2u\left(-\alpha-k^{2}\right)+2k\left(v-2uk\right)\,,\\
\gamma&=&-\left(v-2uk\right)^{2}
+\left(-\alpha-k^{2}\right)\left(1-u^{2}\right)\,,
\end{eqnarray}
where the parameters $v$ and $\Omega$ are related by
\begin{equation}
vk=\left(k^{2}+\alpha\right)u+h+\Omega\,.
\label{par}
\end{equation}
The additional roots $u_{\pm}$ of $F(p)$ are given by 
\begin{eqnarray}
u_{\pm}-u&=&-\frac{\beta}{2\alpha}
\pm\sqrt{\frac{\beta^2}{4\alpha^2}-\frac{\gamma}{\alpha}}\nonumber\\
&=& -u+\frac{uk^2}{\alpha}-\frac{vk}{\alpha}
\pm\sqrt{\frac{-\alpha-\left(v-uk\right)^{2}}{\alpha^{2}}
\left(-\alpha-k^{2}\right)}
\end{eqnarray}

\subsection{The case $\alpha<0$}

\begin{figure}
\begin{center}
\epsfysize=8cm
\epsffile{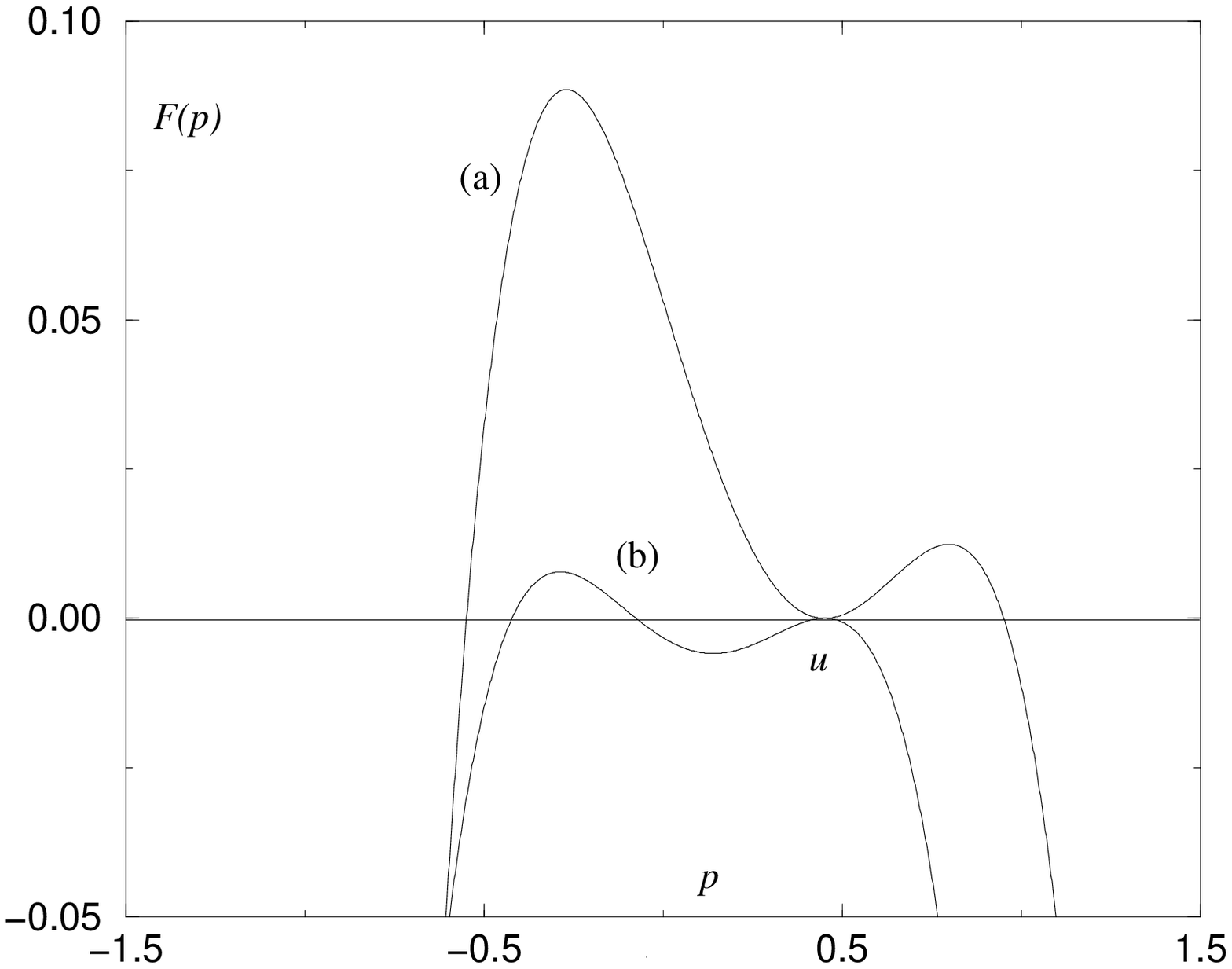}
\caption{The typical behaviour of the polynomial $F(p)$.\label{fig1}}
\end{center}
\end{figure}

\begin{figure}
\begin{center}
\epsfysize=8cm
\epsffile{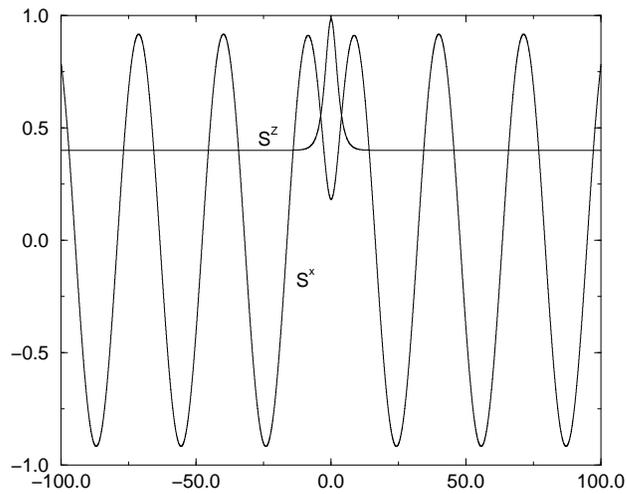}
\caption{Spin field configuration of a $(+)$--soliton with system
parameters $\alpha=-0.5$,~$h=0.1$,~$u=0.4$,~$k=0.2$ and soliton parameters
$v=0.4$,~$x_{0}=0$,~$q_{0}=0$ as a function of $x$ at time $t=0$.\label{fig2}}
\end{center}
\end{figure}

\begin{figure}
\begin{center}
\epsfysize=8cm
\epsffile{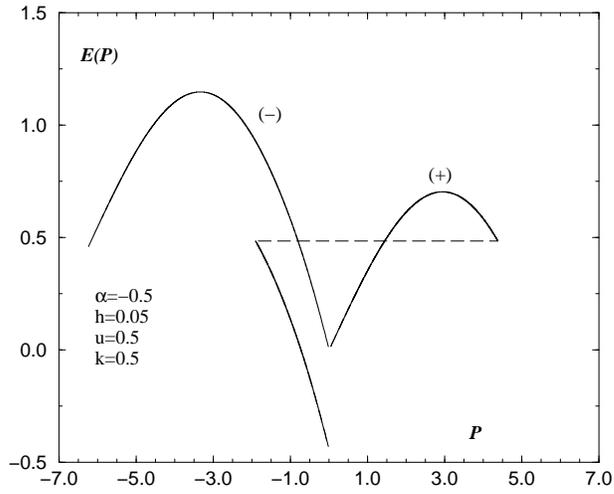}
\caption{Dispersion law $E(P)$ for particular values of the system parameters.
The branches of $(+)$-- and $(-)$--solutions meet in the origin.
The dashed line is a guide to the eye.\label{fig3}}
\end{center}
\end{figure}

The possible behavior of $F(p)$ for $\alpha<0$ and real $u_{\pm}$ is shown 
in figure \ref{fig1}. For a localized solution we need 
$u_{+}\geq u\geq u_{-}$ (a), while otherwise, e.~g. case (b), one obtains a 
periodic wave train oscillating between $u_{+}$ and $u_{-}$. 
If $\alpha<0$ a necessary and sufficient condition for 
$u_{+}\geq u\geq u_{-}$ is $\gamma\geq 0$, which
also ensures the reality of $u_{\pm}$. Thus we have
\begin{equation}
\gamma=-\left(v-2uk\right)^{2}
+\left(-\alpha-k^{2}\right)\left(1-u^{2}\right)\geq 0
\label{width}
\end{equation}
as a restriction to the soliton parameter $v$. Moreover, the desired
solution is only meaningful if it holds $|p(x)|\leq 1$ for all $x$,
i.~e. $|u_{\pm}|\leq 1$. Using $\alpha<0$ one can derive from the
above equations that $u_{+}\leq 1$ is equivalent to
$(v-k(1+u))^2\geq 0$ and $u_{-}\geq -1$ is equivalent to
$(v-k(1-u))^2\geq 0$. Thus, for every choice of the soliton parameter
$v$ compatible with (\ref{width}) we have two solitary solutions
which can be obtained explicitly by elementary integration of
eqs.~(\ref{eom5}),~(\ref{eom3}):
\begin{eqnarray}
p_{\pm}(x,t)&=&u+\frac{\gamma\left(u_{\pm}-u\right)}
{\left(\gamma+\frac{\beta}{2}\left(u_{\pm}-u\right)\right)
\cosh\left(\sqrt{\gamma}\left(x-vt-x_{0}\right)\right)
-\frac{\beta}{2}\left(u_{\pm}-u\right)}\,,
\label{sol1}\\
q_{\pm}(x,t)&=&q_{0}+kx-\omega t\nonumber\\
& &-\tan^{-1}\left(
\frac{v+k\left(1-u\right)}{\sqrt{\gamma}}\frac{u_{\pm}-u}{1+u_{\pm}}
\tanh\left(\frac{\sqrt{\gamma}}{2}\left(x-vt-x_{0}\right)\right)\right)
\nonumber\\
& &-\tan^{-1}\left(
\frac{v-k\left(1+u\right)}{\sqrt{\gamma}}\frac{u_{\pm}-u}{1-u_{\pm}}
\tanh\left(\frac{\sqrt{\gamma}}{2}\left(x-vt-x_{0}\right)\right)\right)
\nonumber\\
\label{sol2}
\end{eqnarray}
The double sign corresponds to a different choice of the integration
constant $p(x_{0},0)=u_{\pm}$, and $q_{0}$, $x_{0}$ are further constants.
The two solutions are related via a formal time inversion, i.~e. the
mapping $(x,t)\mapsto(x,-t)$, $(q,p)\mapsto(q,-p)$,
$(h,\alpha)\mapsto(-h,\alpha)$, $(k,u)\mapsto(k,-u)$,
$(v,\Omega)\mapsto(-v,-\Omega)$. As a consequence we have
$(\beta,\gamma)\mapsto(-\beta,\gamma)$, $u_{\pm}\mapsto-u_{\mp}$,
and the $(+)$-- and the $(-)$-- solution interchange. 
Note also that the Hamiltonian and the equations of motion keep their
form under such operation.\\
We have found the general solitary wave of the form (\ref{ansatz}) obeying 
the boundary conditions (\ref{bc}). The solutions for $p$ are pulse solitons 
with amplitude $(u_{\pm}-u)$ and width $1/\sqrt{\gamma}$ which are 
parametrized by the velocity $v$;  a typical example is given in 
figure \ref{fig2}. From (\ref{sol1}),~(\ref{sol2}) one explitly sees 
the structure of a spin wave (\ref{sw}) for $|x|\gg 1/\sqrt{\gamma}$. 
In this sense these solutions may be called dark solitons.
The phase velocity of the spin wave and 
the soliton velocity differ by $\Omega/k$ (cf.~(\ref{dis}),~(\ref{par})),
which can alternatively be used as soliton parameter. Note also that
the range of admissible velocities given by (\ref{width}) is centered
around the formal group velocity of the spin wave. Moreover, as seen from
eq.~(\ref{mgv}) the maximum (minimum) soliton velocity is given by the minimum
(maximum) group velocity of the linear magnons on the upper (lower) magnon
branch.\\
The above results are in agreement with earlier work by other authors,
beginning with Akhiezer and Borovik \cite{AkBo:67}, who took
the absolute ground state of the model as boundary condition.\\
For $|u|<1$ the inequality (\ref{width}) can only be fullfilled if
\begin{equation} 
-\alpha-k^{2}\geq 0\,.
\label{ex} 
\end{equation}
This is identical with
the condition (\ref{stab}) for the stability of the spin waves.
Thus, the above pair of one--soliton--solutions exists exactly in
the region of boundary parameters $u$,~$k$ where the asymptotic structure
of the solutions is found to be stable.\\ 
If $v$ takes its extremal values, i.~e. $\gamma=0$, and assuming
$\beta>0$,  the $(-)$--soliton
becomes a simple spin wave, while the $(+)$--soliton gets an algebraic
structure:
\begin{eqnarray}
p_{+}(x,t)&=&u+\frac{\frac{\beta}{-\alpha}}
{1+\frac{\beta^2}{-4\alpha}(x-vt-x_{0})^2}
\,,\label{sol3}\\
q_{+}(x,t)&=&q_{0}+kx-\omega t\nonumber\\
& &-\tan^{-1}\left(
\frac{\beta\left(v+k\left(1-u\right)\right)}{-\alpha+\beta-\alpha u}
\frac{\left(x-vt-x_{0}\right)}{2}\right)
\nonumber\\
& &-\tan^{-1}\left(
\frac{\beta\left(v-k\left(1+u\right)\right)}{-\alpha-\beta+\alpha u}
\frac{\left(x-vt-x_{0}\right)}{2}\right)
\label{sol4}
\end{eqnarray}
If $\beta<0$ for an extremal value of $v$ the above results hold vice versa.
An algebraic form of a soliton in such limiting cases has also been 
obtained by Ivanov {\it et al.} \cite{IKM:80} considering the special case
$u=\frac{h}{-\alpha}$,~$k=0$.\\
The energy density of the solitary waves is given by
\begin{equation}
\varepsilon_{\pm}=-\alpha\left(p_{\pm}-u\right)^{2}
+\left(-k\left(v-uk\right)-h-\alpha u\right)\left(p_{\pm}-u\right)\,,
\end{equation}
where we have subtracted the homogenous energy density of the underlying
spin wave. From the explicit solutions (\ref{sol1}),~(\ref{sol2}) we
calculate the total energy and the quantities $P$,~$M$:
\begin{eqnarray}
E_{\pm}&=&2\sqrt{\gamma}-\frac{2h}{\sqrt{-\alpha}}
\left(\tan^{-1}\left(\frac{\beta}{2\sqrt{-\alpha\gamma}}\right)
\pm\frac{\pi}{2}\right)\,,
\label{erg1}\\
P_{\pm}&=&\int dx\left(p_{\pm}\left(\partial_{x}q_{\pm}
\right)-uk\right)\nonumber\\
&=&\left(\tan^{-1}\left(
\frac{\frac{\beta}{2}\left(1+u\right)-\gamma}
{\sqrt{\gamma}\left(v+k\left(1-u\right)\right)}\right)
\pm\frac{\pi}{2}{\rm sign}\,\left(v+k\left(1-u\right)\right)\right)\nonumber\\
& &-\left(\tan^{-1}\left(
\frac{\frac{\beta}{2}\left(1-u\right)+\gamma}
{\sqrt{\gamma}\left(v-k\left(1+u\right)\right)}\right)
\pm\frac{\pi}{2}{\rm sign}\,\left(v-k\left(1+u\right)\right)\right)\,,
\label{mom1}\\
M_{\pm}&=&\int dx\left(p_{\pm}-u\right)\nonumber\\
&=&\frac{2}{\sqrt{-\alpha}}
\left(\tan^{-1}\left(\frac{\beta}{2\sqrt{-\alpha\gamma}}\right)
\pm\frac{\pi}{2}\right)\,.
\label{ang1}
\end{eqnarray} 
In eqs.~(\ref{mom1}),~(\ref{ang1}) we have also subtracted the
homogenous contributions of the nonlinear spin wave from the integrands 
in (\ref{mom0}),~(\ref{ang0}).
For vanishing magnetic field and a given velocity both solutions have the 
same energy. In particular, for $h=0$,~$\beta>0$ the limiting $(+)$--soliton
(\ref{sol3}),~(\ref{sol4}) has the same energy as the $(-)$--soliton, i.~e.
the homogenous spin wave. Eq.~(\ref{erg1}),~(\ref{mom1}) provide a parametric
representation of the dispersion law $E(P)$ in terms of the soliton velocity.
In figure \ref{fig3} we give an example for the function $E(P)$ for both
soliton types.

\subsection{The case $\alpha\geq 0$}

For $\alpha>0$ and $|u|<1$ we have shown that the spin wave solutions
are unstable. Moreover, as the integration of eq.~(\ref{eom5}) is 
formally the same as for $\alpha<0$, we see from (\ref{sol1}) that
we need $\gamma>0$ for a localized solution. Because of (\ref{ex}) this 
cannot be achieved for nonnegative $\alpha$. Thus no localized solution
of the form (\ref{ansatz}) exists for $|u|<1$\\
The case $|u|=1$ is particular. The corresponding spin wave realizes
the absolute minimum of $\cal H$ and the wavenumber $k$ becomes
irrelevant. For such boundary condition well--known solitons exist,
which can be derived similarly by elementary integration
(Kosevich {\it et al.} \cite{KIK:77}, Long and Bishop
\cite{LoBi:79}). These solitons are characterized by two independent
parameters $v$ and $\Omega$ because the relation (\ref{par}) is not valid
if the boundary condition (\ref{eom6}) is not imposed. Therefore
the soliton can alternatively be described by two independent conserved
quantities, the total momentum and the
total angular momentum, and the exact dipersion law $E=E(P,M)$ can be
obtained \cite{KIK:77,Sas:82}.
These two--parametric solutions are also valid for $\alpha\leq 0$ and 
suffciently large $(h+\Omega)$, and under further conditions algebraic 
solitons arise as limiting cases \cite{IKM:80}.

\section{Stability and scattering of solitons}

\begin{figure}
\begin{center}
\epsfysize=10cm
\epsffile{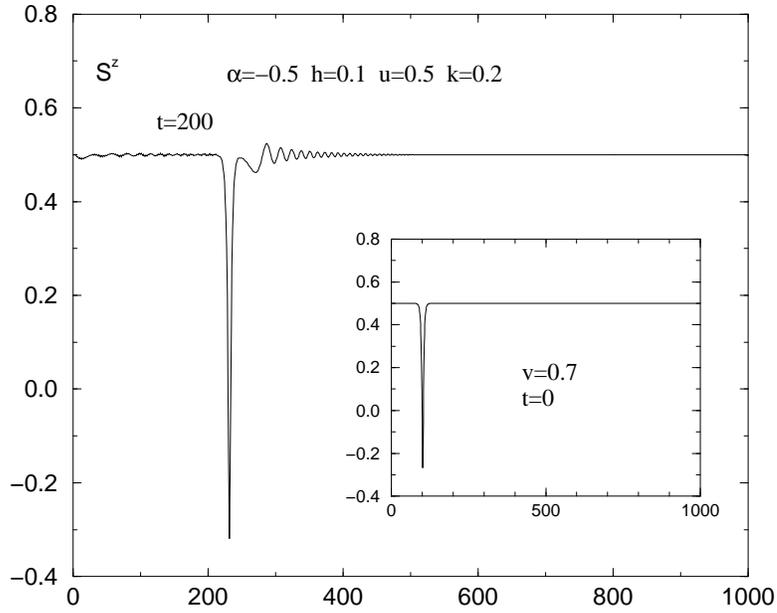}
\end{center}
\caption{Time evolution with a continuum solitary wave as initial data
(small graph, $(-)$--solution) on a lattice of $1000$ sites.
For this choice of parameters comparatively large magnons are emitted in the 
first few hundred time units;
only the component $S^z$ is shown.\label{fig4}}
\end{figure}

\begin{figure}
\begin{center}
\epsfysize=10cm
\epsffile{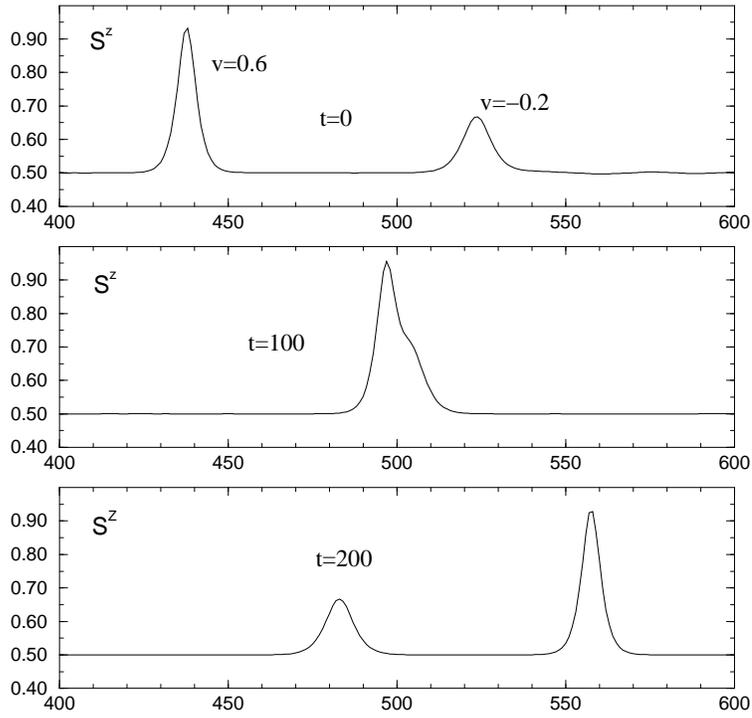}
\end{center}
\caption{Soliton scattering in a system with 
$\alpha=-0.5$,~$h=0.2$,~$u=0.5$,~$k=0.25$; only the component $S^z$ is shown.
\label{fig5}}
\end{figure}

\begin{figure}
\begin{center}
\epsfysize=8cm
\epsffile{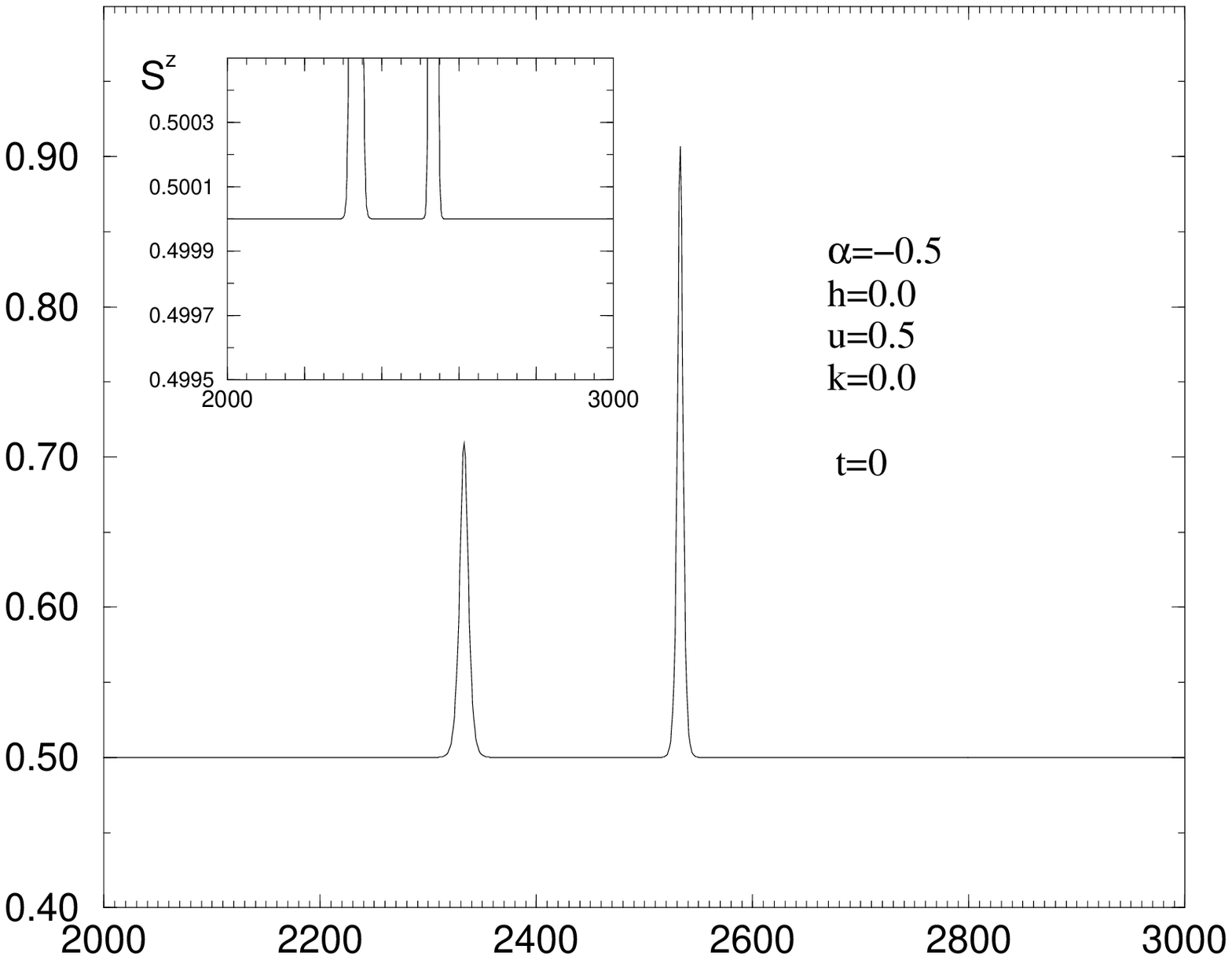}
\epsfysize=8cm
\epsffile{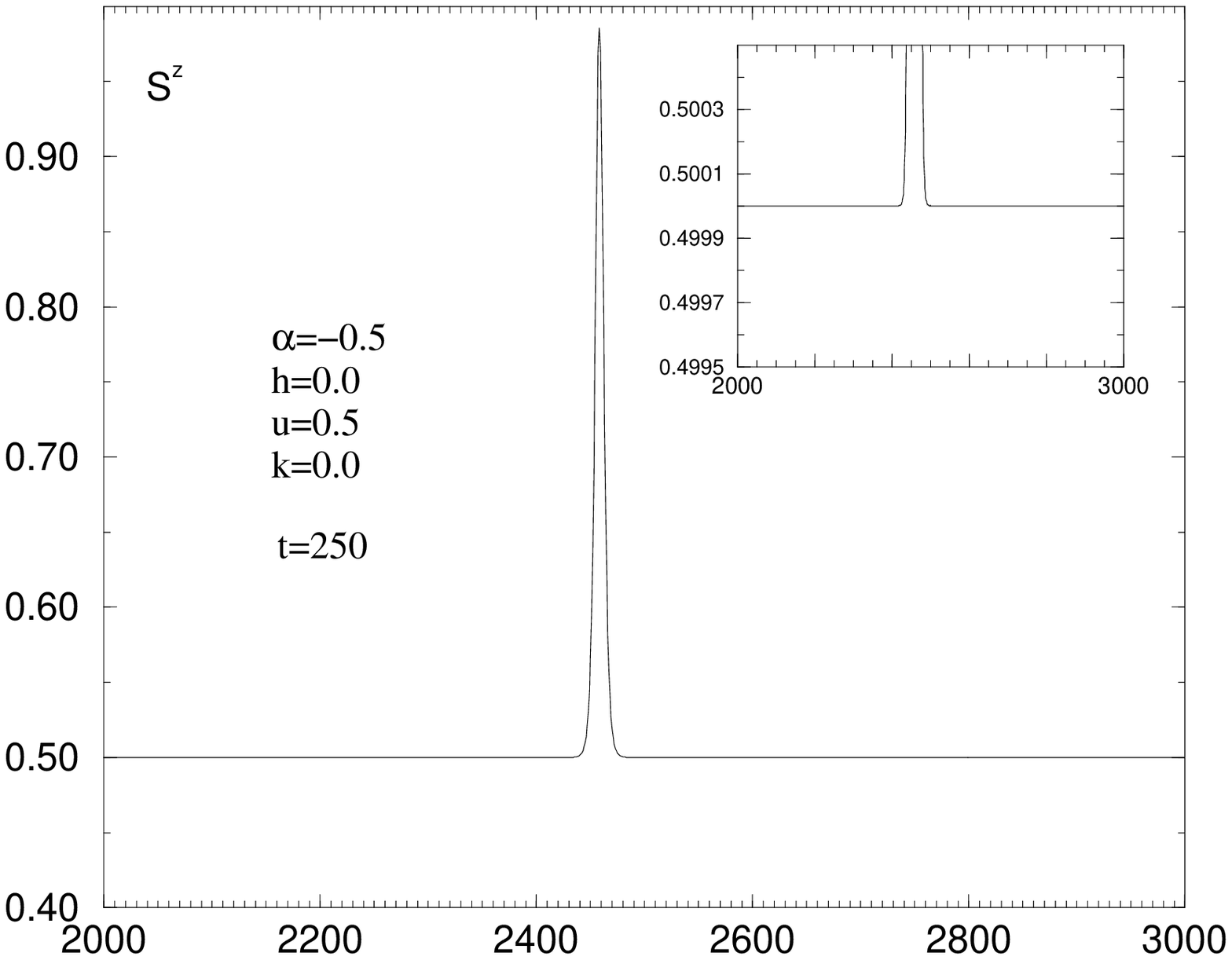}
\epsfysize=8cm
\epsffile{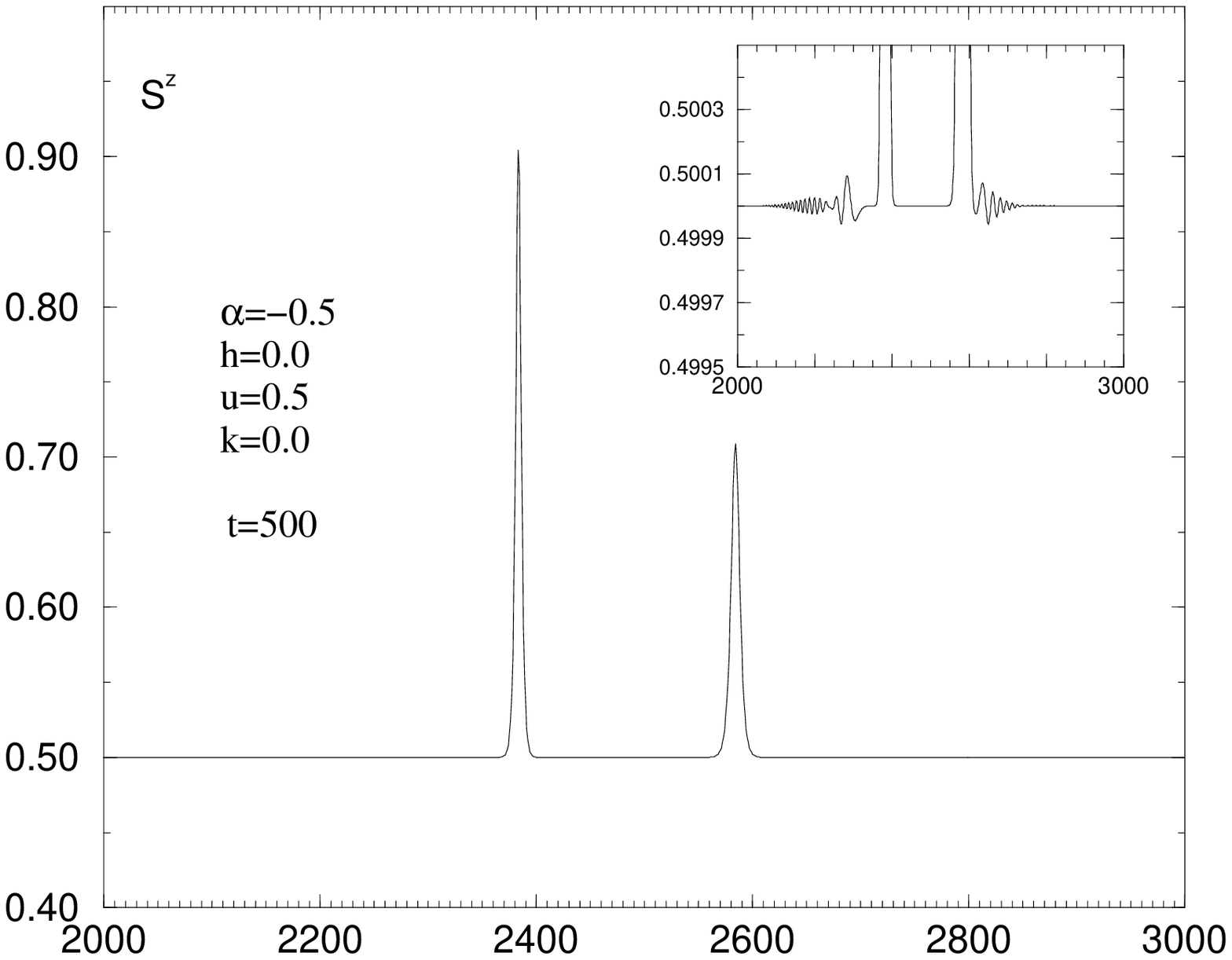}
\end{center}
\caption{A scattering experiment with very pure initial conditions.
In the insets the $z$--component of the spins is plotted with a magnified scale
showing the emission of very small magnon packets as a result of the
scattering process.\label{fig6}}
\end{figure}

We now want to establish the stabitlity of the solitary waves under interaction
with magnons and under scattering with each other. The stability is clear if 
one can prove the integrability of the model for the generalized boundary
conditions used here. The solitary waves have some remarkable
similarity to dark solitons occuring as solutions to the Nonlinear
Schr\"odinger equation with repulsive interaction \cite{Rem:96}.
The latter evolution equation is well--known to be integrable \cite{FaTa:87}. 
Moreover, Nakamura and Sasada \cite{NaSa:82} have proposed a gauge 
transformation mapping the Nonlinear Schr\"odinger equation with attractive 
(repulsive) interaction on the Heisenberg chain with easy--axis (easy--plane)
anisotropy, where the boundary conditions to the spin models correspond to
the absolute ground state. Unfortunately, for the easy--plane case which is
of interest here, this work suffers from some errors pointed out by other
authors \cite{KuPa:83,Kot:84}. Thus, we shall use other means to examine the
question of stability.\\
In order to provide a brief demonstration of the physical significance of
the solitary waves presented before and also to study discreteness
effects to these continuum solutions we have carried out numerical simulation
for the discrete model (\ref{defmoda}).
The spin dynamics is given by the following Landau--Lifshitz--equation: 
\begin{equation}
\partial_{t}\vec S_{n}=\vec S_{n}\times\left(\vec S_{n-1}+\vec S_{n+1}\right)
+h\left(\vec S_{n}\times\vec e_{z}\right)
+\alpha \left(\vec S_{n}\cdot\vec e_{z}\right)
\left(\vec S_{n}\times\vec e_{z}\right)
\label{LL}
\end{equation}
with $\vec S_{n}^{2}=1$ and $\vec e_{z}$ being the unit vector in 
$z$--direction. This equation may also be written in an explicitly 
Hamiltonian form, e.~g. using the canonical conjugate variables
$p_{n}=\cos\vartheta_{n}$,~$q_{n}=\varphi_{n}$. It is seen easily that
the discretized version of the nonlinear spin wave, i.~e. 
$p_{n}=u$,~$q_{n}=kn+\omega t+{\rm const.}$, is an exact solution of 
the dicrete model \cite{RoTh:88}. The natural time unit 
of the system is chosen to be unity by the definition (\ref{defmoda}).
We numerically integrated the equations of motion (\ref{LL}) on lattices of 
up to 5000 sites using a Runge--Kutta scheme of fourth order with a 
stepwidth of $0.01\dots 0.05$. The boundary conditions were implemented
by rotating the spins on the terminal lattice sites with a constant frequency
$\omega$ given by (\ref{dis}) while the $z$--component has the constant
value $u$.\\
If one uses the one--soliton solution of the continuum model evaluated
at the discrete lattice sites as initial data for a simulation, magnons
are emitted from the soliton during the first few hundred time units.
This is an discreteness effect since the continuum solitary wave can only
be an approximate solution to the discrete model. After this process the
soliton moves with constant velocity over the lattice, with no measurable sign
of instability over several thousand time units.
The separation of magnons and solitons corresponds to the result in the
continuum model that the admissible soliton velocities and magnon group
velocities are strictly separated. In figure \ref{fig4} we show an 
example where this effect is comparatively strong for a certain choice of
parameters.
Moreover, small magnons are also radiated from the edges of the system during
the simulation.\\ 
Since the spin configuration outside a soliton decreases rapidly to the 
underlying nonlinear spin wave, different one--soliton solutions can be
matched and scattering experiments can be carried out. If one uses 
appropriately matched continuum solitons as initial data, the two solitons
can apparently pass each other keeping perfectly their shape and identity.
More precisely, we found no effect of instability on the background of magnons 
that stem from the adjustment to the discrete system in the beginning of the
simulation or are radiated from the edges. This holds for all values of 
the system parameters and energy of both $(+)$-- and $(-)$--solitons. A typical
example is shown in figure \ref{fig5}.\\ 
For more precise measurements the magnon background has to be reduced. 
This can be done by using the soliton shape after the adjustment to the
lattice, i.~e. the emission of magnons, as initial data for further 
simulations. Different spin configurations obtained by this procedure can
be matched for a scattering experiment. This method leads to much purer 
initial conditions and works particularly well in the case $h=0$,~$k=0$,
where the spin configuration outside the solitons is spatially 
constant. In order to prevent the small magnons from the edges 
disturbing the scattering process one can choose a sufficiently large system.
In figure \ref{fig6} we show a typical scattering process.
On the natural scale of the problem the two solitons apparently pass each 
other without any sign of instability. If the scale of the vertical axis
is magnified by a factor of about $10^{3}$ one clearly sees that 
the solitons do not perfectly keep their shape but magnons with 
amplitudes of order $10^{-4}$ are emitted as a result of the scattering 
process. We have repeated this experiment with different stepwidths of the 
Runge--Kutta scheme and confirmed that this effect is not a 
numerical artifact.\\
This observation strongly indicates the nonintegrability of the discrete
model in the strict mathematical sense. Of course, this does not imply any
statement about the integrability of the continuum model under the generalized
boundary conditions. In particular not,
since the discrete model shows properties very near to those of integrable 
models.\\
In summary, we have demonstrated by numerical simulations that our solitary 
waves are physically significant objects showing a highly distinct but not
absolute stability in their time evolution. 

\section{Conclusions}

In this work we have examined the classical Heisenberg chain with
uniaxial anisotropy under generalized boundary conditions.
For an anisotropy of easy--plane type two branches of solitons are found
on the background of a nonlinear spin wave that is used as boundary condition.
The parameter area of stability of the nonlinear spin wave solution coincides
with the existence of the solitary waves.
For the easy--axis and the isotropic model we have shown that no further 
one--soliton solutions exist apart from those already known. 
Moreover, in these cases the spin wave solutions are found to be unstable.
This fact has apparently not been realized yet, since the spin wave solutions 
(\ref{sw}) have been discussed for such systems by various authors
without mentioning their instability \cite{MiSt:91,KIK:90,Fog:80,LRT:76}.\\
For the case $\alpha<0$ we have examined the stability of the continuum
solitary waves by numerical simulations of the discrete Heisenberg chain.
The solitary waves turn out to be  highly but not absolutely stable in
their time evolution and especially under scattering with each other.
Thus the discrete model is shown to have properties very near to those
typical for integrable models, but small effects indicating nonintegrability
are clearly observed. From these numerical results and also from the 
similarity of the solitary waves obtained here to dark solitons of 
the Nonlinear Schr\"odinger equation one can strongly conjecture 
the integrability of the continuum model.\\
The spin wave state (\ref{sw}) could serve as a phenomenological model 
for a one--dimensional easy--plane magnet at finite (not necessarily very low) 
temperature.
In thermal equilibrium the spatially averaged energy density and 
magnetization have certain values determining the parameters $u$,~$k$. 
However, the long--range ordering in this one--dimensional system must be 
destroyed at finite temperature by thermal fluctuations \cite{MeWa:66}. 
The linear and localized nonlinear excitations obtained in this work provide
a possible mechanism for this effect.


\newpage

\newpage

\newpage

\newpage

\end{document}